\documentclass[]{article}

\usepackage{amsmath}
\usepackage{graphicx}
\usepackage{cite}
\usepackage{srcltx}
\usepackage{color}

\begin{document}

\title{Entropy production\\ in continuous phase space systems}
\author{David Luposchainsky and Haye Hinrichsen\\[2mm]
\small Universit\"at W\"urzburg, Fakult\"at f\"ur  Physik und Astronomie\\
\small 97074  W\"urzburg, Germany}

\maketitle

% this is prescribed by JSP guidelines
\setlength{\textwidth}{27pc}
\setlength{\textheight}{43pc}

\def\d{{\rm d}}
\def\0{\emptyset}
\def\Stot{\ensuremath{S_{\text{tot}}}}
\def\Ssys{\ensuremath{S_{\text{sys}}}}
\def\Senv{\ensuremath{S_{\text{env}}}}
\def\localdSenv{\ensuremath{\d S_{\text{env}}^{\text{loc}}}}  % local entropy production after integration
\renewcommand\vec[1]{\mathbf #1}
\def\xvec{\vec{x}}
\def\d{{\rm  d}}

\def\del#1{\frac{\partial}{\partial#1}}

\newcommand\Todo[1]{{\color{red}[TODO: #1]}}

\begin{abstract}
We propose an alternative method to compute the environmental entropy production of a classical underdamped nonequilibrium system, not necessarily in detailed balance, in a continuous phase space. It is based on the idea that the Hamiltonian orbits of the corresponding isolated system can be regarded as microstates and that entropy is generated in the environment whenever the system moves from one microstate to another. This approach has the advantage that it is not necessary to distinguish between even and odd-parity variables. We show that the method leads to a different expression for the differential entropy production along an infinitesimal stochastic path. However, when integrating over all possible paths the local entropy production turns out to be the same as in previous studies. This demonstrates that the differential entropy production in continuous phase space systems is not uniquely defined.
\end{abstract}

%\pacs{}

\parskip 2mm

%=================================================
\section{Introduction}
%=================================================

In the past decade an important advance of nonequilibrium statistical physics has been the development of stochastic thermodynamics~\cite{SeifertReview,SpinneyFordReview}. In this approach thermodynamic quantities such as the entropy are defined as functionals along the microscopic stochastic trajectory of the system in its configuration space, allowing one to study fluctuations around the average. This led to the discovery of various fluctuation theorems which generalize the second law of thermodynamics~\cite{Lebowitz,Gallavotti,Gaspard,Seifert2005}. For example, the total entropy of a system together with its environment in a nonequilibrium steady state is known to obey the integral fluctuation theorem $\langle e^{-\Delta\Stot} \rangle =1$, implying the second law $\langle \Delta \Stot \rangle \geq0$.

The concept of stochastic thermodynamics is most easily introduced in the context of stochastic Markov jump processes. A Markov jump process is defined by a space of discrete classical configurations $c \in \Omega$ and certain transition rates $w_{c\to c'}(t)\geq 0$ for spontaneous jumps from $c$ to $c'$. As time evolves, such a system moves randomly in its own configuration space, producing a stochastic trajectory
\begin{equation}
\gamma: \; c_0\to c_1 \to c_2\to\ldots
\end{equation}
of instantaneous transitions taking place at certain transition times $t=t_1,t_2,\ldots$. 

In stochastic thermodynamics it is assumed that the total entropy change can be written as a sum of the changes of the internal entropy of the system and the entropy of the environment~\cite{SeifertReview}
\begin{equation}
\Delta\Stot(t)=\Delta\Ssys(t)+\Delta\Senv(t).
\end{equation}
Here the internal entropy $\Ssys(t)$ is a fluctuating quantity defined by
\begin{equation}
\label{SystemEntropy}
\Ssys(t) = -\ln p_{c(t)}(t)\,,
\end{equation}
where $p_c(t)$ denotes the probability to find the system in the configuration $c$ at time $t$. Since this probability evolves deterministically according to the master equation 
\begin{equation}
\label{MasterEquation}
\dot p_c(t)=\sum_{c'\in\Omega} \bigl[p_{c'}(t)w_{c'\to c}(t)-p_c(t)w_{c \to c'}(t)\bigr]\,
\end{equation}
the internal entropy $\Ssys(t)$ evolves smoothly superposed by discontinuous jumps whenever the system hops to a different configuration. Averaging $\Ssys(t)$ over the probability distribution $p_c(t)$, one retrieves the usual Boltzmann-Gibbs or Shannon entropy
\begin{equation}
\langle \Ssys(t) \rangle = -\sum_c p_c(t) \ln p_c(t) ~.
\end{equation}
The second contribution $\Delta \Senv$ comes from the fact that nonequilibrium systems are usually driven from outside, i.e. they interact with the environment (reservoir). On the level of the Markov process the environment is usually not modeled explicitly, rather it is implemented implicitly by means of asymmetric rates ($w_{c'\to c}\neq w_{c \to c'}$). For example, a system describing the flow of heat between two reservoirs at different temperatures is usually described in terms of a biased diffusion process which drives the particles on average in one direction. In a nonequilibrium steady state the internal entropy $\langle \Ssys(t) \rangle$ of such a system will be on average constant, but the incessant external drive will continually increase the entropy in the environment. 

Remarkably, the external entropy production~$\Delta \Senv(t)$ can be quantified even if the physical properties of the environment are not known. In fact, as shown in~\cite{Schnakenberg,Gaspard}, whenever the system hops from $c$ to $c'$, the environment entropy changes discontinuously by the amount
\begin{equation}
\label{DiscreteEntropyProduction}
\Delta \Senv(t) = \ln \frac{w_{c\to c'}(t)}{w_{c'\to c}(t)}\,,
\end{equation}
irrespective of the physical realization of the environment.

The aim of the present paper is to study the external entropy production $\Delta \Senv$ in systems with \textit{continuous} degrees of freedom. For such systems the probability density evolves according to the Fokker-Planck equation, replacing the master equation~(\ref{MasterEquation}). As will be explained below, the corresponding entropy production is usually determined by computing the logarithmic ratio of the statistical weights of a stochastic path and an appropriate conjugate or reversed path, analogous to Eq.~(\ref{DiscreteEntropyProduction}) in the discrete case. This approach is safely established for \textit{overdamped} systems, where the microstate of a particle is given by its position~\cite{Seifert2005}.

For \textit{underdamped} systems with momenta, on which we will focus in the present work, the situation is less clear. The reason is that is is not so obvious how the path has to be reversed in order to obtain a meaningful expression for the entropy production. One possibility is to change the sign of all momenta under path reversal, which requires to distinguish even and odd state variables. Based on this approach, Spinney and Ford (SF) recently derived a general expression for the differential entropy production of continuous systems in full phase space~\cite{SpinneyFordContinuous}. However, some aspects of their differential entropy production appear to be physically implausible, which motivated us to study this problem. For example, in the simple special case of a particle in equilibrium with a heat bath, the entropy production is expected to be related to the dissipated heat
\begin{equation}
T\, \d \Senv = - \d Q = - \d\left[ \frac{p^2}{2m} + V(q) \right] = - \frac p m \d p - V'(q) \, \d q\,,
\label{eqn:standard-thermo-result}
\end{equation}
where the minus sign comes from the fact a positive heat $\d Q>0$ flows away from the reservoir. Surprisingly, as we will see below in Eq.~(\ref{RestingParticleSF}), the corresponding result obtained by Spinney and Ford turns out to be different.

This discrepancy can be traced back to the fact that it is not clear how to identify a `microstate' or `configuration' of a continuous stochastic underdamped system in the framework of classical mechanics. The approach used by SF implicitly assumes that the microstates are just the points $(q,p)$ in phase space. Here we propose a different viewpoint, suggesting that the microstates should be identified with the underlying Hamiltonian orbits one would obtain without friction and heat bath. Using this approach the reversed path is constructed forward in time without changing the sign of the momenta, meaning that it is no longer necessary to distinguish even and odd variables. We arrive at a different expression for the differential entropy production $\d\Senv(q,p,\d q,\d p,\d t)$ which is physically more plausible and no longer in contradiction with the special case in Eq.~(\ref{eqn:standard-thermo-result}).

However, the differential entropy production $\d\Senv(q,p,\d q,\d p,\d t)$ along in infinitesimal path is only an intermediate result. As we are ultimately interested in the entropy production during an infinitesimal time span $\d t$ at a certain point in phase space, the differential entropy production still has to be integrated over all infinitesimal paths weighted by their probability. As a result one obtains the \textit{local} differential entropy production $\localdSenv(q,p,\d t)$ which depends on the phase space coordinates only. Surprisingly our method and the one used by SF give the same expression for the local entropy production, meaning that both approaches are correct and equally legitimate. 

This leads us to the main conclusion that the differential entropy production $\d\Senv(q,p,\d q,\d p,\d t)$ along an infinitesimal stochastic path is to some extent ambiguous, a circumstance that to our knowledge has not been noted before. This ambiguity is in some sense analogous to the one occurring in stochastic calculus, where different integration schemes (It\={o}/Stratonovich) with suitable drift terms may lead to the same result.

The paper is organized as follows. In the next section we introduce basic definitions and notations. In Sect. 3 we summarize the approach by SF, discuss the meaning of points in phase space, suggest to consider Hamiltonian orbits instead, and compute the corresponding expression for the differential entropy production. Finally, in Sect. 4 we show in the example of a underdamped particle that the integration over all trajectories leads to the same result in both cases, testing the special case of thermal equilibrium.

%=================================================
\section{Definitions and Notations}
%=================================================

\paragraph{Overdamped continuous systems:}
In the overdamped limit the time scale of interest is much larger than the correlation time of the momenta which allows one to model them as an effective noise. For example, the overdamped dynamics of a single particle with the coordinate $q(t)$ is given by the Langevin equation
\begin{equation}
\dot q = \beta D F(q,t) + \zeta(t)\,,
\end{equation}
where $F(q,t)$ is a force, $\zeta(t)$ is a white Gaussian noise with temporal correlations $\langle \zeta(t)\zeta(t')\rangle=2D\delta(t-t')$, $D$ is the diffusion constant, and $\beta D$ is the mobility of the particle.

The entropy production for overdamped Langevin systems in a continuous state space is defined in the same spirit as in the discrete case: Monitoring the system up to time $T$, the particle moves along a stochastic path $\gamma=\{q(t)\}$ starting at position $q_0$ which occurs with a certain probabilistic weight $\mathcal{P}[\gamma|q_0]$. The entropy production in the environment along this path is given by~\cite{Seifert2005}
\begin{equation}
\Delta \Senv = \ln \frac{P[\gamma|q_0]}{P^\dagger[\gamma^\dagger|q_0^\dagger]}\,,
\end{equation}
where the symbol $\dagger$ denotes the operation of path reversal $q^\dagger(t)=q(T-t)$, combined with the exchange of final and initial position $q_0^\dagger = q_T$ and the reversal of the protocol $t\to T-t$ for time-dependent forces (corresponding to time-dependent rates in the discrete case). Obviously, this eqpression is just a continuum version of Eq.~(\ref{DiscreteEntropyProduction}), where the positions $q$ play the same role as the configurations $c$ in the discrete case. 

It is important to note that the operation $\dagger$ does not involve a \textit{physical} time reversal operation, i.e. there is nothing ``running backward in time''. In fact, $\gamma^\dagger$ is just the reflected path and $P^\dagger[\gamma^\dagger|q_0^\dagger]$ its statistical weight in the \textit{same} physical process running forward in time, only using a reflected protocol in the case of time-dependent transition rates.

\paragraph{Underdamped continuous systems:}
In the \textit{underdamped} case, where the particle is characterized by positions $q$ and momenta $p$, the situation is more complicated. If we reversed the stochastic trajectory naively by replacing
\begin{equation}
\gamma=\{q(t),p(t)\} \quad\longrightarrow\quad \gamma_{\rm naive}^\dagger=\{q^\dagger(t),p^\dagger(t)\}=\{q(T-t),p(T-t)\} 
\end{equation}
we would obtain a conjugate trajectory running backward in space but with the velocity pointing in the wrong direction. This problem can be overcome either by adding an explicit reflection of time $t \to -t$, which means $\gamma^\dagger$ is actually running backward in time, or -- formally equivalent -- by redefining the $\dagger$-operation in such a way that it changes the sign of $p$ by itself. Using the latter approach Spinney and Ford (SF) introduced a general formalism that distinguishes between odd and even variables under the $\dagger$-operation~\cite{SpinneyFordPRL}, i.e.
\begin{equation}
\gamma=\{q(t),p(t)\} \quad\longrightarrow\quad \gamma^\dagger=\{q^\dagger(t),p^\dagger(t)\}=\{\epsilon q(T-t),\epsilon p(T-t)\} 
\end{equation}
with $\epsilon = +1$ or $-1$ for even and odd quantities such as $q$ and $p$ respectively. With this formalism Spinney and Ford were able to compute the entropy production of arbitrary Langevin systems with inertia, including \textit{underdamped} equations of motion in the full $(q,p)$ phase space [cf.  Eqs. (74) and (86) of Ref.~\cite{SpinneyFordContinuous}].

%=================================================
\begin{figure}
	\centering
	\includegraphics[width=0.7 \linewidth]{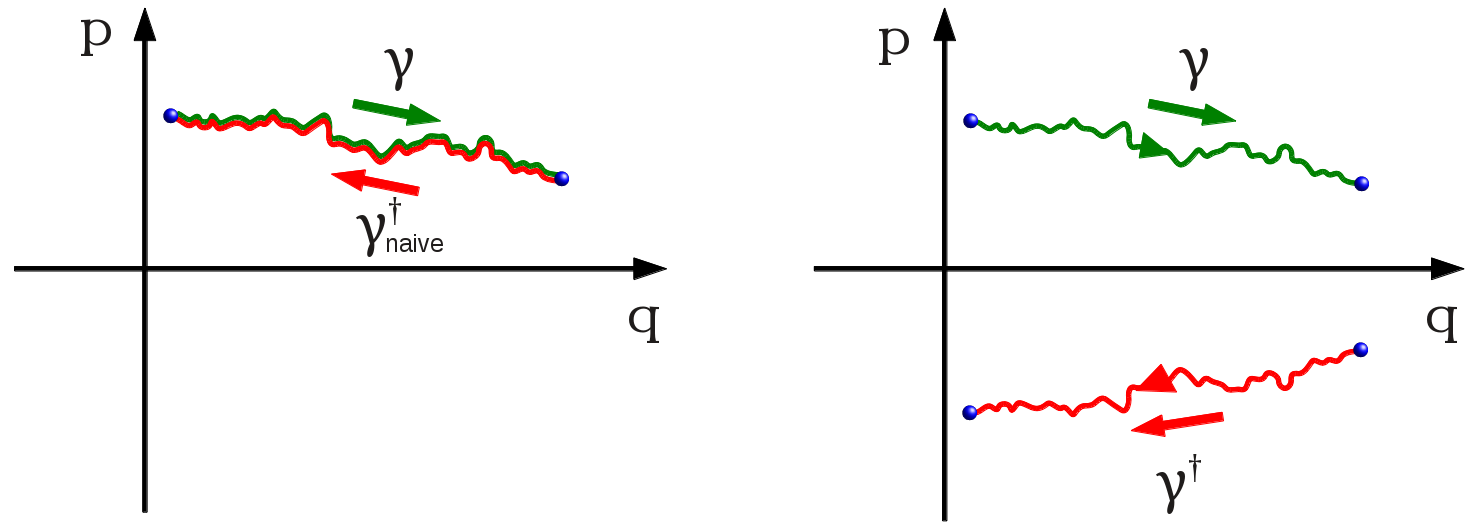}
	\caption{\footnotesize [Color online] Path reversal in phase space. Left: Reversing the trajectory naively by replacing $t \to T-t$ leads to an unphysical path where momentum and velocity are oriented in opposite direction. Right: Path conjugation proposed by Spinney and Ford, flipping in addition the sign of the momentum.}
	\label{fig:reversal}
\end{figure}
%=================================================

Nevertheless it remains unsatisfactory that the redefined $\dagger$-operation maps the trajectory $\gamma$ to a completely different part of phase space, where the form of the orbits may differ from the reflected ones (see Fig.~\ref{fig:reversal}). As for the terminal points of the trajectory, instead of simply exchanging $q_0$ with $q_T$, one now substitutes $(q_0,p_0)\to (q_T,-p_T)$ and $(q_T,p_T)\to (q_0,-p_0)$, i.e. the two terminal points are replaced with a \textit{different} pair of points, which seems to be inconsistent with the spirit of Eq.~(\ref{DiscreteEntropyProduction}). We believe that this circumstance plays a role in the context of the so-called \textit{odd-parity variable problem} which is currently discussed in the community~\cite{OddParity}. 

In the present paper we demonstrate that the strict distinction between odd and even variables is not needed in order to define a physically meaningful entropy production. Following the lines of Seiferts approach in Ref.~\cite{Seifert2005} we define the differential entropy production by a logarithmic ratio of the weights of two forward processes, identifying the underlying Hamiltonian orbits as the microstates of the system. This approach is conceptually simpler as it is no longer necessary to distinguish between even and odd variables. Moreover, it is possible to consider Langevin equations with arbitrary functions which are neither symmetric nor antisymmetric. As already pointed out in the Introduction, we find that the differential entropy production $\d\Senv(q,p,\d q,\d p,\d t)$ differs from the one computed by using Spinney and Ford's approach. However, integrating over all possible trajectories one arrives at the same result, meaning that the two approaches are equivalent. Therefore, we are led the conclusion that the differential entropy production is to some extent ambiguous, meaning that different expressions can describe the same physical situation.

\paragraph{Model studied in this work:}
In what follows we study a particle with phase space coordinates $\bigl(q(t),p(t)\bigr)$ in a potential $V(q)$ which is in contact with an external heat bath. The stochastic trajectory of the particle is described by the Langevin equation
\begin{equation}
\label{Langevin}
\dot q = p\,, \qquad \dot p = -V'(q) - \mu(p) p + \Gamma(p)\xi\,.
\end{equation}
As usual $\xi(t)$ denotes a white Gaussian noise with correlations $\langle \xi(t) \xi(t') \rangle = 2D\delta(t-t')$ which is understood to be integrated in the It\={o} sense throughout the whole paper. For simplicity we set the mass of the particle to $m=1$. In order to include the possibility of nonlinear friction we assume the coefficient $\mu(p)$ and the noise amplitude $\Gamma(p)$ to be $p$-dependent functions. In the simplest case of linear friction the coefficients $\mu$ and $\Gamma$ are constant and obey the Einstein relation $\Gamma=\sqrt{2 T \mu}$ in thermal equilibrium.

\paragraph{Fokker Planck equation:}
%--------------------------------------
Following the notation of Ref.~\cite{SpinneyFordContinuous} and defining phase space vectors $\xvec = \bigl(q,p\bigr)$ the Langevin equation (\ref{Langevin}) can be written as a stoachastic differential equation of the form
\begin{equation}
\d x_i = A_i(\xvec,t)\d t+B_i(\xvec,t)\d W_i\,,
\end{equation}
where $\d W_i$ denotes the Wiener process generated by the noise $\xi(t)$. For It\={o} integration, the probability distribution $P(\xvec,t)$ to find the particle in the phase space point $\xvec = \bigl(q,p\bigr)$ at time $t$ evolves according to the Fokker Planck equation
\begin{equation}
\label{FP}
\partial_t P(\xvec,t) = -\sum_i \partial_{x_i} J_i(\xvec,t)\,,
\end{equation}
where 
\begin{equation}
J_i(\xvec,t)=A_i(\xvec,t)P(\xvec,t)-\partial_{x_i}[D_i(\xvec,t)P(\xvec,t)]
\end{equation}
is the dynamical probability current with the diffusion coefficients
\begin{equation}
D_i(\xvec,t)=\frac12 B_i(\xvec,t)^2 \,.
\end{equation}
More specifically, for the Langevin equation~(\ref{Langevin}) describing a single particle in a potential $V(q)$ the Fokker-Planck equation (\ref{FP}) is given by
\begin{equation}
\begin{split}
\label{LangevinQP}
\partial_t P(q,p,t)
&=
\Bigl(\bigl[\mu (p)+p \mu'(p)+\Gamma (p) \Gamma''(p)+\Gamma '(p)^2\bigr]
\\&\qquad
+ \bigl[p \mu (p)+2 \Gamma (p) \Gamma '(p)+V'(q)\bigr]\partial_p
\\&\qquad
+ \dfrac12{\Gamma (p)^2}\partial^2_p-p\partial_q \Bigr) P(q,p,t) \,.
\end{split}
\end{equation}

\paragraph{Generalized Einstein relation:}
%-----------------------------------------
For a single particle in a heat bath the Langevin equations~(\ref{Langevin}) can be split into a component given by the Hamilton equations of motion with the Hamiltonian $\mathcal H(q,p)=p^2/2 + V(q)$ and a component $- \mu(p) p + \Gamma(p)\xi$ accounting for the heat bath. Of particular interest are dynamical rules for which the system evolves into a stationary Boltzmann-Gibbs distribution of the form
\begin{equation}
\label{BoltzmannState}
p_{\rm eq}(q,p) = \frac{e^{-\beta \mathcal H(q,p)}}{Z} = \frac{1}{Z} \exp\left[-\beta\Bigl(\frac{p^2}{2}+V(q)\Bigr)\right]\,,
\end{equation}
where $\beta=1/T$ and $k_B := 1$. Inserting this distribution on the r.h.s. of the Fokker-Planck equation one obtains a first-order differential equation in $\mu(p)$ with the solution
\begin{equation}
\mu(p) \;=\; \frac12 \beta \Gamma(p)^2 - \frac{\Gamma(p)\Gamma'(p)}{p} + \frac{C\,e^{\frac{\beta p^2}2}}{p}\,,
\end{equation}
where $C$ is an integration constant. Since the last term would lead to a divergent mean acceleration of the particle in the stationary state, it is unphysical so that we have to set $C=0$. The remaining expression
\begin{equation}
	\label{detailed_balance}
	2 p \mu(p) \;=\; \beta  p \Gamma (p)^2-2 \Gamma (p) \Gamma '(p)
\end{equation}
relates the noise amplitude $\Gamma(p)$ with the friction coefficient $\mu(p)$, generalizing the Einstein relation $2 \mu=\beta \Gamma^2$ for the special case of linear dissipation and additive noise.

\paragraph{Short-time propagator:}
%-----------------------------------------
An important function, that will be needed to compute the differential entropy production in the following section, is the short-time propagator $G(\xvec'|\xvec; \d t)$ of the Fokker-Planck equation. The short-time propagator can be understood as the probability density to find a particle starting from $\xvec=(q,p)$ at the point $\xvec'$ after an infinitesimal time span $\d t$. 

Since the short-time propagator has to solve the Fokker Planck equation only to lowest order in $\d t$ it is not uniquely defined~\cite{ShortTimeProp1,ShortTimeProp2}. The origin of this ambiguity is two-fold. On the one hand, the \textit{form} of the propagator function does not need to be Gaussian because the central limit theorem guarantees that any distribution with the correct first and second moment will lead to the same macroscopic propagator. On the other hand, there is an additional freedom in choosing the \textit{evaluation point} of the amplitudes $A_i(\xvec,t)$  and $D_i(\xvec,t)$: These functions do not need to be evaluated at the starting point $\xvec$ of the propagator but they could also be evaluated at the final point $\xvec'$ or at any point in the immediate neighborhood of these points. Of course, this would generate corrections which have to be compensated to lowest order by suitable counterterms in the propagator. Although this freedom reminds one of the It\={o}-Stratonovich 
dilemma we emphasize that this ambiguity is an additional freedom which is completely independent of our choice to use the It\={o} scheme in the Langevin equation.

Following Ref. \cite{SpinneyFordContinuous} we will consider a subspace of all possible solutions by introducing a parameter $a \in [0,1]$ and defining the short-time propagator by
\begin{align}
\label{propagator}
G_a(\xvec'|\xvec; \d t) =&\prod_i(4\pi D_i\d t)^{-1/2} \times \\
&\exp\Bigl[ -\,\frac{(\d x_i-A_i \d t+2 a  D'_i\d t)^2}{4 D_i \d t} - a A'_i \d t +a^2 D''_i \d t \Bigr]\,, \nonumber
\end{align}
where the functions and their partial derivatives
\begin{align*}
&A_i=A_i(\vec r,t)\,,\quad A'_i=\partial_{r_i} A_i(\vec r,t)\,,\\
& D_i=D_i(\vec r,t)\,,\quad D'_i=\partial_{r_i} D_i(\vec r,t)\,,\quad D''_i=\partial^2_{r_i} D_i(\vec r,t)
\end{align*}
are evaluated at the point $\vec r=(1-a)\xvec + a\xvec'$ on a straight line between the two points $\xvec$ and $\xvec'$. It is straight forward to verify that this propagator solves the Fokker-Planck equation to first order in $\d t$.

\paragraph{Short-time propagator for an underdamped particle:}
%-----------------------------------------
The underdamped particle in Eq.~(\ref{Langevin}) is described by a system of two differential equations, namely, a deterministic equation $\dot q=p$ and a stochastic one for $\dot p$. To compute the short-time propagator we first add an infinitesimally small noise in the first component, i.e. we consider the Langevin equations%
\begin{equation}
\label{LangevinEpsilon}
\dot q = p + \epsilon \, \xi_q\,, \qquad \dot p = -V'(q) - \mu(p) p + \Gamma(p)\xi_p\,
\end{equation}
with two uncorrelated Gaussian noise functions $\xi_q(t)$ and $\xi_p(t)$. Inserting this into Eq.~(\ref{propagator}) one obtains
\begin{align}
\label{specificG}
G^\epsilon_a(q',p'|q,p;\, \d t) =& 
\frac{1}{2\pi \epsilon \Gamma(r) \d t}\,\exp\biggl[
a^2  \Bigl(\Gamma (r) \Gamma ''(r)+\Gamma '(r)^2\Bigr) \d t \\&
-\frac{\Bigl(  V'(q + a \d q)\d t+2 a   \Gamma (r) \Gamma '(r) \d t +\d p +
   r \gamma (r)\d t \Bigr)^2}{2  \Gamma (r)^2 \d t }\notag\\
   &+a   \left(r \mu'(r)+\mu(r)\right)\d t-\frac{(\d q - r\, \d t )^2}{2  \epsilon ^2\, \d t }\notag
\biggr]\,,
\end{align}
where $\d q=q'-q$, $\d p=p'-p$, and 
\begin{equation}
\label{rdef}
r=r(p,p')=p+a (p'-p) 
\end{equation}
is the momentum at which the functions are evaluated. As expected, by taking $\epsilon\to 0$ we obtain
\begin{align}
\label{Gdelta}
G_a(q',p'|q,p;\, \d t) =& 
\frac{\delta(\d q-r\,\d t)}{\sqrt{2\pi \, \d t}\; \Gamma(r)}\,\exp\biggl[
a^2  \Bigl(\Gamma (r) \Gamma ''(r)+\Gamma '(r)^2\Bigr) \d t \\&
-\frac{\Bigl(  V'(a \d q +q)\d t+2 a   \Gamma (r) \Gamma '(r) \d t +\d p +
   r \mu(r)\d t \Bigr)^2}{2  \Gamma (r)^2 \d t }\notag\\
   &+a   \left(r \mu '(r)+\mu (r)\right)\d t\notag
\biggr]\,
\end{align}
which is deterministic in the position coordinate.

%=================================================
\section{Differential entropy production}
%=================================================

The differential entropy $\d\Senv$ can be understood as a Taylor expansion of $\Delta\Senv$ to lowest order in $\d q$, $\d p$ and $\d t$ along an infinitesimal line element of the trajectory $\gamma$ in phase space. In the following we show that this differential entropy production can be defined in different ways.

\paragraph{Differential entropy production according to Spinney and Ford:}
%-------------------------------------------------------------------------
As already outlined in the Introduction, the amount of entropy generated in the environment along a stochastic trajectory $\gamma$ is given by
\begin{equation}
\label{ContinuousEntropyProduction}
\Delta \Senv = \ln \frac{P[\gamma|\xvec_0]}{P^\dagger[\gamma^\dagger|\xvec_0^\dagger]}.
\end{equation}
where $\dagger$ stands for a suitable path conjugation operation. As suggested by Spinney and Ford~\cite{SpinneyFordPRL}, this conjugation has to change the sign of odd-parity variables such as momenta, i.e.
\begin{equation}
\gamma: \; (q_0,p_0) \to (q_T,p_T) \qquad \stackrel\dagger\longrightarrow \qquad \gamma^\dagger: \; (q_T,-p_T) \to (q_0,-p_0) \,.
\end{equation}
Using this definition of path reversal, the differential entropy production according to Spinney and Ford (see Eq.~(25) of Ref.~\cite{SpinneyFordContinuous}) along an infinitesimal section of a path is given by
\begin{equation}
\label{SFEP}
\d \Senv^{\rm SF} =  \ln\frac{G_a(\xvec'|\xvec;\d t)}{G_{b}({\xvec'}^\dagger|\xvec^\dagger;\d t)}\,,
\end{equation}
where $a,b$ are free parameters reflecting the ambiguity of the short-time propagator. More specifically, in the special case of the underdamped particle Eq.~(\ref{LangevinEpsilon}) we have
\begin{equation}
\label{SFEPUD}
\d \Senv^{\rm SF} = \lim_{\epsilon\to 0} \ln\frac{G^\epsilon_a\bigl(q',p'\, | \, q,p\, ; \d t\bigr)}
{G^\epsilon_{b}\bigl(q,-p \, | \, q',-p'\,;\,\d t\bigr)}\,.
\end{equation}
Note that the differential $\d \Senv^{\rm SF}$ is generally inexact since it cannot be derived from a corresponding thermodynamic potential.

Clearly, the limit $\epsilon\to 0$ can only be carried out if the two propagators peak at the same position in phase space. Roughly speaking this means that the two $\delta$-functions (see Eq. (\ref{Gdelta})) have to cancel out when taking the limit individually in the propagators since otherwise the entropy production would be locally divergent. This leads directly to the condition
\begin{equation}
(q'-q)-\Bigl(p+a(p'-p)\Bigr)\d t \;=\; -\Bigl[(q-q')-\Bigl(-p'+b(-p+p')\Bigr)\d t\Bigr]
\end{equation}
with the solution $b=1-a$, replacing a lengthy derivation in the appendix A of Ref.~\cite{SpinneyFordContinuous}. Setting $a=0$ and $b=1$ and using Spinney and Ford's approach, the differential entropy production can be expanded to first order in $\d t$ and to second order in $\d p$, arriving at
\begin{equation}
\begin{split}
\label{SFEntropyProduction}
\d \Senv^{\rm SF} &=\frac{1}{\Gamma^2}
\Bigl(
-\mu  \Gamma ^2-\Gamma ^3 \Gamma ''+\Gamma ^2 {\Gamma '}^2-\Gamma ^2 p \mu '+2 \mu  \Gamma  p \Gamma '-2 \mu  p V'-2 \Gamma 
   \Gamma ' V'
\Bigr)\,\d t\\
&+  \frac{1}{\Gamma^2}  
\Bigl(-2 \Gamma  \Gamma '-2 \mu  p\Bigr) \,\d p
 \;\;+ \mathcal O (\d t^2) +  \mathcal O (\d p^3) \,,
\end{split}
\end{equation}
where for the sake of brevity we dropped the arguments of the functions $\Gamma(p),\mu(p)$, and $V(q)$ which are all evaluated at the point $(q,p)$.

As already pointed out in the Introduction, this expression has some physically implausible properties. For example, in the simplest case of a particle subjected to linear friction  ($\mu(p) = \mu = \text{const}$) in thermal equilibrium with a heat bath ($\Gamma=\sqrt{2 T \mu}$) the formula given above reduces to
\begin{equation}
\label{RestingParticleSF}
\d \Senv^{\rm SF} \;=\; -\beta \,p\,\d p- \,\beta V'(q) \, \d q - \mu \, \d t 
\end{equation}
which differs from the usual textbook result~(\ref{eqn:standard-thermo-result}) by the last term $-\mu \d t$. This extra term would imply that a free particle ($V=0$) resting at some point in space ($p=0$) continually produces negative differential entropy in the environment -- a surprising result since in the discontinuous case a system resting in a particular configuration does not produce any entropy. 

\paragraph{On the meaning of points in phase space:}
%---------------------------------------------------
In the arguments given above and in most papers about entropy production in continuous systems, it is implicitly assumed that the configurations $c$ in the  discrete case should somehow correspond to individual points in phase space, i.e. the points in phase space themselves are identified as the `microstates' of the system. However, as we will argue in the following, it is not clear whether this is the only meaningful way to introduce the notion of microstates.

To see this, let us disconnect the external reservoir (in our example by setting $\mu(p)=\Gamma(p)=0$). What is left over is a classical Hamiltonian system which, being disconnected, does not produce entropy in the environment. The same happens in a discrete system which stays in a single configuration $c$. However, unlike the latter, the disconnected Hamiltonian system does not stay in a single point, instead it still moves along its deterministic trajectory governed by the Hamiltonian equations of motion. This leads us to the conjecture that all points \textit{connected by the Hamiltonian flow} should be considered as belonging to the same `microstate' of the system. In other words, we suggest that the `microstates' of the system are the underlying Hamiltonian orbits rather than individual points in phase space.

\begin{figure}
	\centering
	\includegraphics[width=0.9 \linewidth]{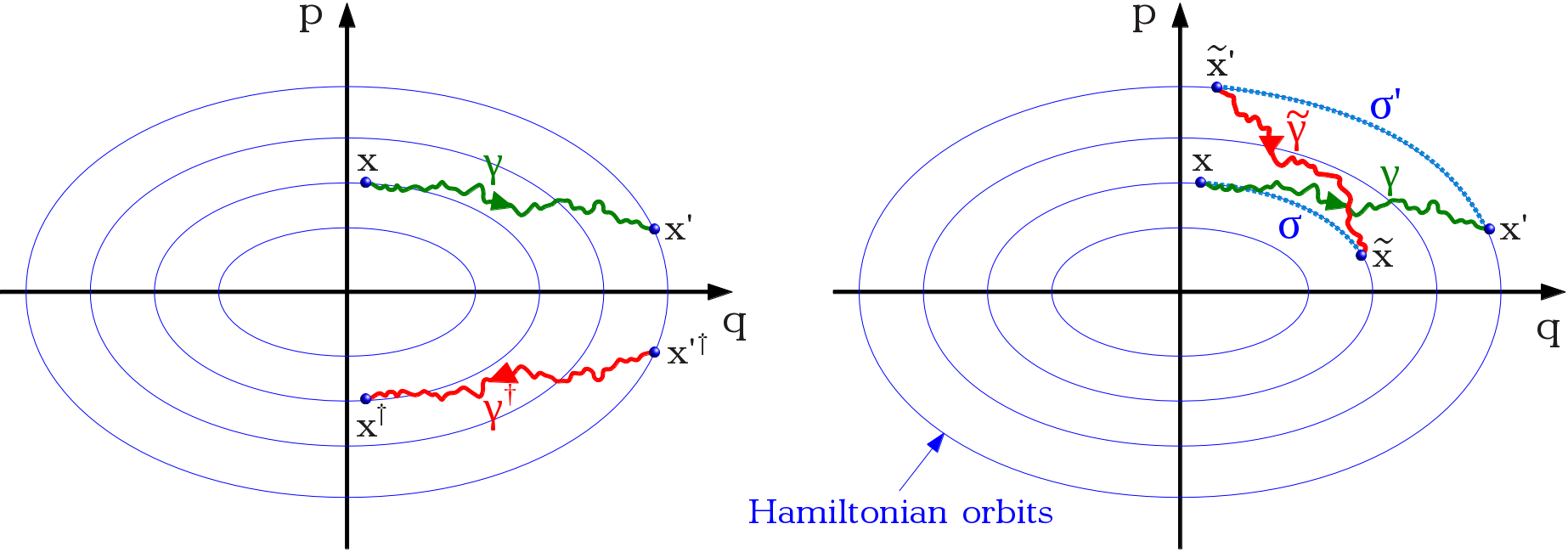}
	\caption{ \footnotesize [Color online] Two different schemes to define the conjugate path. Left: Spinney and Ford's approach mirrors the terminal points at the \(q\)-axis into a different part of phase space. Right: In our approach the terminal points are instead exchanged along the underlying Hamiltonian orbits $\sigma$ and $\sigma'$.}
	\label{fig:trajectories}
\end{figure}

Reconnecting the system with the external reservoir, the thermal noise will randomly drive the system away from one Hamiltonian orbit to another. We suggest that such a thermally induced jumps from one Hamiltonian orbit to another can be interpreted as a spontaneous transitions to a different `microstate', analogous to a change of the configuration $c \to c'$ in the discrete case. This would imply that only transitions between different Hamiltonian orbits should change the entropy in the environment, opposed to the formula (\ref{fig:trajectories}) suggested by Spinney and Ford.

\paragraph{Differential entropy production based on Hamiltonian orbits:}
%----------------------------------------------------------------------------
Based on this idea we propose a new definition of environmental entropy production with respect to the Hamiltonian orbits instead of points in phase space. This definition does not require parity operations, meaning that it is not necessary to distinguish between odd and even variables. As we will see, this leads in fact to a different expression for the differential entropy production. However, after integration we will be led to exactly the same physical predictions as Spinney and Ford, suggesting that the definition of the differential entropy production is not unique.

Our method is applicable to systems which can be separated into a Hamiltonian part and another contribution coming from the external environment so that it is possible to identify the underlying Hamiltonian orbits. 

Our approach is motivated as follows: As shown in Fig.~\ref{fig:trajectories}, a given stochastic path $\gamma$ typically connects two different underlying Hamiltonian orbits $\sigma$ and $\sigma'$. According to the arguments given above, the conjugate path, which is denoted here by $\tilde\gamma$ (in order to distinguish it from $\gamma^\dagger$ used by Spinney and Ford), should run from $\sigma'$ to $\sigma$ in such a way that the terminal points of the trajectories are crosswise connected by the Hamiltonian flow (see Fig.~\ref{fig:trajectories}). This ensures that both $\gamma$ and $\tilde\gamma$ are defined as forward trajectories in the \textit{same} part of phase space so that the special treatment of odd variables is no longer necessary:

\begin{center}
\begin{tabular}{|c|c|c|c|} \hline
path & transition & starting point & ending point \\ \hline
$\gamma$ & $\sigma\to\sigma'$ & $\xvec=(q,p)$ & $\xvec'=(q',p')$ \\
$\tilde\gamma$ & $\sigma'\to\sigma$ & $\tilde\xvec=(\tilde q,\tilde p)$ & $\tilde\xvec'=(\tilde q',\tilde p')$ \\
\hline
\end{tabular}
\end{center}
Using this conjugation scheme, we define the differential entropy production in the spirit of Eq.~(\ref{DiscreteEntropyProduction}) as
\begin{equation}
\d\Senv^{\text{HF}}(\xvec'|\xvec\,;\,\d t) = \ln\frac{G_a(\xvec '|\xvec \,; \d t)}{G_{b}(\tilde\xvec|\tilde\xvec' \,;\d t)}\,,
\end{equation}
where the phase space points $\xvec$ and $\tilde \xvec$ (and likewise $\tilde \xvec'$ and $\xvec'$) are connected by the Hamiltonian flow (HF) over the time span $\d t$. The parameters $a,b$ reflect the ambiguity of the short-time propagator and will be determined later. Moreover, we dropped a possible explicit time dependency of the propagator.

\paragraph{Application to a single particle:}
%----------------------------------------------------------------------------
In the example of a single underdamped particle, we have
\begin{align}
\tilde \xvec  &= \bigl(\tilde q , \tilde p\bigr) = \bigl(q + p \d t,\,p + f(q) \d t\bigr) \\
\tilde \xvec' &= \bigl(\tilde q' , \tilde p'\bigr) = \bigl(q'-p' \d t,\, p'- f(q') \d t\bigr)
\end{align}
so that
\begin{equation}
\d\Senv = \lim_{\epsilon\to 0}\ln\frac{G^\epsilon_a\bigl(q',p'\,|\,q,p\, ;\,\, \d t  \bigr)}
{G^\epsilon_{b}\bigl(q+p \d t,\, p+f(q)\d t\,\,|\,\, q'-p'\d t, \,p'-f(q') \d t  \, ;\,\, \d t \bigr)}\,.
\end{equation}
Again this expression can only be evaluated if the singularities arising in the limit $\epsilon\to 0$ cancel out, leading to the condition
\begin{align}
&(q'-q)-\Bigl(p+a(p'-p)\Bigr)\d t \\
=&
-\Bigl[(q-q')+(p+p')\d t-\Bigl(p'-f(q')\d t+b\bigl(p-p'+(f(q)+f(q'))\d t\bigr)\Bigr)\Bigr]\notag
\end{align}
Expanding the force by $f(q')=f(q)+f'(q)(q'-q)+\mathcal O(\d t^2)$ and neglecting terms of the order $\d t^3$ this equation is solved for $a=b=1/2$. This result is very plausible: It simply means that the functions appearing in the two propagators have to be evaluated at the same point in phase space. In the case considered by Spinney and Ford, apart from the reflection in the momenta, this means that the position coordinates have to coincide, giving $a=b$. In our method, on the other hand, this happens only at the particular point where the two trajectories cross each other, namely, roughly in the middle of the four points $(\xvec,\xvec',\tilde\xvec,\tilde\xvec')$.

Inserting $a=b=1/2$ the differential entropy production suggested by us is given by
\begin{align}
\label{HF}
\d\Senv^{\text{HF}} \;&=\; \frac{1}{\Gamma^2} \Bigl(      -2 p \mu  V'-2 \Gamma  \Gamma ' V' \Bigr)\d t \;+\;
           \frac{1}{\Gamma^2} \Bigl( -2  p \mu -2\Gamma  \Gamma ' \Bigr) \d p\\
        &\;\;+ \frac{1}{\Gamma^2} \Bigl( -p \mu '+\frac{2 p \mu  \Gamma '}{\Gamma }-\mu -\Gamma  \Gamma ''+\Gamma '^2 \Bigr) \d p^2 \notag
        \;+\mathcal O(\d t^2) + \mathcal O(\d t^3)\,,
\end{align}
which is our main result. Again we dropped the arguments of the functions $\Gamma(p),\mu(p)$, and $V(q)$ which are all evaluated at the point $(q,p)$. Obviously, this differential entropy production differs from the one  proposed by Spinney and Ford in Eq.~(\ref{SFEntropyProduction}) in various respects:
\begin{itemize}
 \item In contrast to their result, our differential entropy production vanishes to first order in $\d t$ along the deterministic trajectory $\d p =-V'(q)\d t$.
 \item While Spinney and Ford's expression is linear in $\d t$ and $\d p$, our version also includes a second-order term $\d p^2$.
 \item In our case the differential entropy production of a particle in a potential with linear friction and noise in thermal equilibrium with a heat bath correctly reproduces the thermodynamically expected result, see Eq.~(\ref{RestingParticleSF}).
\end{itemize}

%=================================================
\section{Local entropy production rate}
%=================================================

\paragraph{Definition:}
%----------------------------------------------------------------------------
%
The \textit{local} differential entropy production $\localdSenv(\xvec)$ is defined as the infinitesimal entropy production averaged over all paths originating at the point $\xvec$ weighted with the corresponding probability of the path:
\begin{equation}
\localdSenv (\xvec; \d t) \;=\; \int\limits_{-\infty}^{+\infty} \d \xvec' \; G_c(\xvec'|\xvec;\,\d t) \; \d \Senv (\xvec'|\xvec;\,\d t) + \mathcal{O}(\d t^2) \,.
\end{equation}
Again this differential is generally inexact. Note that $c\in[0,1]$ is another free parameter which reflects the ambiguity in the short-time propagator which is independent of the parameters $a,b$ appearing in the differential entropy production and also independent of the integration scheme used in the Langevin equation. 

In the example of a single particle we have
\begin{equation}
\label{LocalEntropyProduciton}
\localdSenv (q,p; \d t) \;=\; \int\limits_{-\infty}^{+\infty} \d q' \int\limits_{-\infty}^{+\infty}\d p' \; G_c(q',p'|q,p;\, \d t) \; \d \Senv (q',p'|q,p;\, \d t) + \mathcal{O}(\d t^2)  \,.
\end{equation}
Since for $\d t \to 0$ the integral is dominated by short paths we can Taylor-expand the differential entropy production by
\begin{equation}
\d \Senv (q',p'|q,p;\, \d t) \;=\; \sum_{i,j=0}^\infty \sigma_{ij}(q,p;\d t)\; (q'-q)^i (p'-p)^j + \mathcal{O}(\d t^2) \,.
\end{equation}
Inserting this expansion back into Eq.~(\ref{LocalEntropyProduciton}) we can rewrite the local entropy production as
\begin{equation}
\label{ShortForm}
\localdSenv (q,p; \d t) \;=\; \sum_{i,j=0}^\infty \sigma_{ij}(q,p;\d t)\; M_{ij;\,c}(q,p;\d t) + \mathcal{O}(\d t^2) 
\end{equation}
with the moments
\begin{equation}
\label{moments}
M_{ij;\, c}(q,p;\d t)
\;=\;
	\int\limits_{-\infty}^{+\infty} \d q'
	\int\limits_{-\infty}^{+\infty} \d p'
	\;
	G_c(q',p'|q,p;\, \d t)
	\;
	(q'-q)^i (p'-p)^j
	\,.
\end{equation}
As we are interested in taking the limit $\d t\to 0$ in Eq.~(\ref{ShortForm}) it is sufficient to compute $\sigma_{ij}(q,p;\d t)$ and $M_{ij}(q,p;\d t)$ only to first order in $\d t$.

\paragraph{Calculation of the moments for an overdamped particle:}
%----------------------------------------------------------------------------
%
Let us now compute the local entropy production rate for the overdamped particle
\begin{equation}
	\begin{split}
		\dot q &= p \\
		\dot p &= -V'(q) - \mu(p)p + \Gamma(p)\xi(t)\,.
	\end{split}
	\label{dynamics}
\end{equation}
We calculated the moments $M_{ij;\,c}(q,p;\d t)$ in Eq.~(\ref{moments}) by standard methods using a two-fold application of the saddle point method combined with a limit $\epsilon\to 0$, finding the following results:
\begin{align}
	M_{00;\,c} &= 1 + \mathcal O(\d t) \\
	M_{01;\,c} &= \Bigl(-p \mu (p) - V'(q)\Bigr) \; \d t + \mathcal O(\d t^2) \\
	M_{02;\,c} &= \Gamma(p)^2 \; \d t  + \mathcal O(\d t^2) \\
	M_{ij;\,c} &= \mathcal O(\d t^{i+1}) \quad (i,j>0)
\end{align}
Remarkably, the parameter $c$ in the short-time propagator in Eq.(\ref{moments}) does not appear in the relevant leading order terms but only in higher-order corrections.

Now the local entropy production rate (\ref{ShortForm}) can be calculated directly, and it turns out that both the approach by Spinney and Ford and our variant yield exactly the \emph{same} result, namely,
\begin{align}
	\begin{split}
		\localdSenv(q,p; \d t)
		&\;= \;\Bigl[
		- \mu(p)
		+ \frac{2 p^2 \mu(p)^2}{\Gamma(p)^2}
		- p \mu'(p)
		\\&\qquad
		+ \frac{4 p \mu(p) \Gamma'(p)}{\Gamma(p)}
		+ \Gamma'(p)^2
		- \Gamma(p) \Gamma''(p)\Bigr] \d t + \mathcal O(\d t^2)\,.
	\end{split}
\end{align}
How is it possible that two different expressions for the differential entropy production lead to the same result after integration? The answer is that our expression includes an additional term of the order $\d p^2$, which does not appear in SF's expression. The terms of order $\d p$ and $\d p^2$ both give a contribution proportional to $\d t$ after integration, hence they may partially compensate each other. In this way the two expressions give the same result to linear order in $\d t$ after integration.

In order to obtain the global entropy production rate averaged over the whole phase space as a function of time, this expression has to be integrated weighted by the actual probability distribution [evolving according to Eq.~(\ref{LangevinQP})]
\begin{equation}
\label{GlobalSDot}
\langle \d\Senv \rangle(t) = \d t \int_{-\infty}^{+\infty}\d q \int_{-\infty}^{+\infty}\d p \; p(q,p,t) \localdSenv(q,p)\,.
\end{equation}

\paragraph{Entropy production in thermal equilibrium:}
%----------------------------------------------------------------------------
%
Let us finally consider the special case of thermal equilibrium with multiplicative noise. In this case the functions $\mu(p)$ and $\Gamma(p)$ obey the generalized Einstein equation (\ref{detailed_balance}), which is synonymous for transition rates obeying detailed balance. Inserting this relation the differential entropy production according to Spinney and Ford's approach and the present paper is given by
\begin{align}
\d \Senv^{\rm SF} &= -\beta p \,\bigl(\d p + V'(q)\d t\bigr) - \frac\beta2 \Gamma(p)^2\d t \ + \mathcal{O}(\d p^3) + \mathcal{O}(\d t^2)\\
\d \Senv^{\rm HF} &= -\beta p \,\bigl(\d p + V'(q)\d t\bigr) + \mathcal{O}(\d p^2) + \mathcal{O}(\d t^2)\,.
\end{align}
As one can see, only the second version vanishes along the deterministic trajectory $\dot p=-V'(q)$, in agreement with the assumptions that Hamiltonian orbits do not produce entropy. Again the local entropy production rate
\begin{align}
	\localdSenv(q,p,\d t)
	&= \left[
	- \frac12 \, \beta \, \Gamma(p)^2
	+ \frac12 \, \beta^2 \, p^2 \Gamma(p)^2
	- \beta \, p \, \Gamma(p) \Gamma'(p)\right] \, \d t + \mathcal O (\d t^2)\,
	\label{sDotDB}
\end{align}
obtained by integration of $p$ and $q$ (including the $\d p^2$ term in Eq.~(\ref{HF})) turns out to be the same in both cases. Note that this quantity may be nonzero even in equilibrium since the entropy production is a fluctuating quantity. However, the mean entropy production rate (\ref{GlobalSDot}) integrated over phase space vanishes in the stationary state (\ref{BoltzmannState}), i.e.
\begin{equation}
\langle\d\Senv\rangle(t)
\;=\;
	\int\limits_{-\infty}^{+\infty}\d q
	\int\limits_{-\infty}^{+\infty}\d p
	\,
	\frac{1}{Z}
	e^{-\beta\big(\frac{p^2}2-V(q)\big)}
	\localdSenv(q,p)
\;=\;
0\,,
\end{equation}
as can be shown by splitting the last term in (\ref{sDotDB}) into two equal halves and then rewriting one of them using integration by parts. This is plausible since for equilibrium system, including those with generalized Einstein relations, the average entropy production rate in the environment should be zero.

%====================================
\section{Conclusion}
%====================================

In this paper we have proposed a definition of the differential entropy production along a stochastic trajectory in continuous phase space systems with inertia. Our approach does not depend on the distinction between even and odd variables, instead it requires to split up the dynamics into an underlying deterministic and a stochastic part. The key idea is to consider a system that moves along its deterministic trajectory as remaining in the same state and that only jumps between different Hamiltonian orbits produce entropy in the environment. In other words, instead of points in phase space we identify the underlying Hamiltonian orbits as the microstates of the system.

Following these physically motivated ideas we arrive at an expression for the differential entropy production $\d \Senv^{\text{HF}}(q,p,\d q,\d p; \d t)$ which differs from a previously studied expression proposed by Spinney and Ford. In particular, it vanishes along the underlying Hamiltonian orbits and correctly reproduces the well-known relation $T \d \Senv = -\d Q$ in the special case of thermal equilibrium. 

The apparent contradiction can be resolved by computing the local entropy production rate $\localdSenv$ which turns out to be the same in both cases. This demonstrates that the notion of differential entropy production incorporates an element of ambiguity, which is resolved in the subsequent step of calculating expectation values over all possible trajectories. As there are various short-time propagators generating the same time evolution, we arrive at the main conclusion that various versions of the differential entropy production generate, when integrated, the same local entropy production rate. A systematical investigation of the full range of this ambiguity would be desirable.

\paragraph{Acknowledgments:\\}
%----------------------------------------------------------------------------

\noindent HH would like to thank U. Seifert and A. C. Barato for interesting discussions and helpful remarks.

%====================================

\end{document}